\documentstyle[twoside,fleqn,espcrc2]{article}
\newcommand{\La}{\Lambda_{QCD}}

\newcommand{\be}{\begin{equation}}
\newcommand{\ee}{\end{equation}}
\newcommand{\bea}{\begin{eqnarray}}
\newcommand{\eea}{\end{eqnarray}}
\newcommand{\as}{\alpha_s}
\newcommand{\aQ}{\alpha_s(Q^2)}
\newcommand{\al}{\alpha}
\newcommand{\la}{\lambda}
\newcommand{\ra}{\rightarrow}
\newcommand{\om}{\omega}
\def\slashchar#1{\setbox0=\hbox{$#1$}           
   \dimen0=\wd0                                 
   \setbox1=\hbox{/} \dimen1=\wd1               
   \ifdim\dimen0>\dimen1                        
      \rlap{\hbox to \dimen0{\hfil/\hfil}}      
      #1                                        
   \else                                        
      \rlap{\hbox to \dimen1{\hfil$#1$\hfil}}   
      /                                         
   \fi}                                         %

\newcommand{\AmS}{{\protect\the\textfont2
  A\kern-.1667em\lower.5ex\hbox{M}\kern-.125emS}}

\title{A Quick Look at Renormalons\thanks{Based on talks at QCD'96, Montpellier,
July '96 and at the Workshop on Renormalons, Copenhagen, August '96.  To appear
in the proceedings of QCD'96.}}

\author{R. Akhoury and V.I. Zakharov\address{Department of Physics, 
        University of Michigan, \\ 
        Ann Arbor, Michigan-48109, USA}}
       
\begin{document}

\begin{abstract}
We present a sketchy review of renormalon-based phenomenology.
In particular, the leading,
$1/Q$ corrections to various observables, KLN cancellations
for power-suppressed corrections and the fixation of operator
matrix elements are highlighted.
 
\end{abstract}

\maketitle
Renormalons as a pure theoretical construct are known since 1977
\cite{thooft}. The first attempts to develop a 
renormalon-based phenomenology
are more recent \cite{mueller,vz}. 
Nowadays it is a fast developing field and a concise 
review is obviously beyond  
the scope of the present contribution. Instead, this paper 
is a compromise between an
original paper and a mini-review. Namely we try to 
put the topics discussed in
original papers with the participation of the authors into a more
general framework.

\section{THEORETICAL ASPECTS.}

\subsection{Renormalon basics.}

The image of renormalons is invariably produced
by the renormalon chain which is an 
insertion of $n$, where $n$ is large, vacuum-polarization
bubbles into a photon (gluon) line. 
Denote, furthermore, by $k$, the 4-momentum flowing through the
dressed line and by $Q$ a large external parameter (like
total energy in $e^+e^-$-annihilation). If we expand in $\alpha (Q^2)$
then in the $n$-th order one readily obtains for the coefficient
$a_n$ in front of $\alpha (Q^2)$ the 
following estimates in terms of the first coefficient of the
$\beta$-function $b_0$:
\be
(a_n)_{IR}~\sim~\int d^4k\left(b_0ln Q^2/k^2\right)^n~\sim 
~n!b_0^n2^{-n}\label{series}
\ee
in case of $k^2\ll Q^2$ and 
\be
(a_n)_{UV}~\sim~\int d^2k\left(b_0ln Q^2/k^2\right)^n
~\sim~n!(-b_0)^n
\ee
in case of $k^2\gg Q^2$.
The behaviour (\ref{series}) is process independant
provided there is a single soft gauge-boson line
$k^2 \ll Q^2$ and $Q$ is a euclidean momentum.
In this way renormalons indicate the 
asymptotical nature of perturbative expansions
and hence bring an uncertainty to perturbative calculations.
Estimating this uncertainty in a standard way
one gets :
\bea
\delta_{IR}~\sim~exp(-2b_0/\alpha_s(Q^2))~\sim~(\La/Q)^4,\\ \nonumber
\delta_{UV}~\sim~exp(-b_0/\alpha_s(Q^2))~\sim~(\La/Q)^2
\eea
where 
$\alpha_s(Q^2)\sim~(b_0ln Q^2/\La^2)^{-1}$.  

Thus, renormalons, although arising within a purely
perturbative framework, realize the idea of dimensional transmutation.
Also, renormalons indicate presence of non-perturbative 
power corrections of the same order in
$(\La/Q)$ which are now
needed to render the theory uniquely determined despite the
uncertainties of perturbative expansions. 
Since renormalons always introduce two different
mass scales, that is, $k^2\gg Q^2$ or $k^2\ll Q^2$
it is natural to invoke operator product expansions
to evaluate their contribution. In case of infrared renormalons
it is the standard OPE, when applicable. In particular,
the series (\ref{series}) above, for $n\gg 1$
can be considered as a perturbative
contribution to the matrix element of 
$\langle 0|\alpha_s (G_{\mu\nu}^a)^2|0\rangle$ \cite{mueller,vz}:
\be
{\langle 0|\as(G_{\mu\nu}^a)^2|0\rangle_{ren}
\over 24\pi Q^4}
=\sum_{\scriptstyle{n~large}}
{3\aQ^{n+1}b_0^n\over 2^{n+1}\pi^2}n!\label{cond}
.\ee
The non-perturbative counterpart was in fact 
introduced first via QCD sum rules
\cite{svz}.
In case of UV renormalons one can utilize \cite{vainshtein} a reverse OPE
which is an expansion in $Q^2/k^2$. 
The use of an OPE allows us to formulate the renormalon
contribution in terms of the running coupling, without direct
use of the renormalon chains. The use of the OPE brings also a
challenge to theory \cite{vainshtein,fsp}.
Namely, it turns out that a single renormalon chain does not dominate
in fact over two and more chains. Thus, there is no closed set
of graphs producing the same $n!$.

In short, the renormalons
are a simple and systematic way to
parametrize the IR contributions to various observables.

\subsection{Limitations of renormalons.}

At the one loop level 
renormalons are not a unique and even not necessarily the simplest way
to probe IR regions perturbatively. Another posibility is an introduction
of finte gluon mass $\la$. The gluon mass was tried as a fit parameter
about 15 years ago \cite{cornwall}. In particular  there is an 
infrared-sensitive perturbative 
gluon condensate \cite{chet}:
\be
\langle 0|\aQ(G_{\mu\nu}^a)^2|0\rangle~=~-{3\as\over\pi^2}\la^4ln\la^2
\label{con}\ee
which is a substitution for the renormalon contribution 
(\ref{cond}) in case of massless
gluons. In recent times, the use of a finite mass $\la\neq 0$
has became very common.
In what follows we shall not always distinguish between one-loop calculations
with finite $\la$ and a single renormalon chain, labeling generically both
techniques as renormalons.

It might be worth emphasizing, 
however, that nowadays the finite gluon mass is 
used mostly not as a fit parameter
but rather as a probe of infrared  region. 
Namely, non-analytical in $\la^2$ terms 
come exclusively from infrared gluons. The power of $\la$ characterizes then
the strength of the IR sensitive contributions.
Generically, the translation of one-loop calculations with finite gluon mass
$\la$ and with IR renormalons looks as follows \cite{bbz}:
\bea
\al a_0 ln\la^2+\al a_1{\sqrt{\la^2}\over Q}+\al a_2
{\la^2ln\la^2\over Q^2}+...\ra\\ \nonumber
b_0 ln\La^2+b_1{\La\over Q}+b_2{\La\over Q^2}+...\label{par}
\eea
where we keep only infrared sensitive contributions
and $a_i,b_i$ are coefficients.

Among the limitations of the renormalon 
technique let us mention the following
points:\\
(i) renormalons respect the symmetries of the Lagrangian and cannot,
for example, produce a nonvanishing quark condensate
$\langle \bar{q}q\rangle \neq 0$.\\
(ii) renormalons are "target-blind", e.g.,
$\langle p|G^2|p\rangle~_{renorm}=
\langle 0|G^2|0\rangle_{renorm}$\\
(iii) renormalons give no direct indication of confinement,
say, of a string configuration.

An interesting problem is brought out
by renormalons \cite{blok} in supersymmetric gluodynamics.
To render the theory supersymmetric one adds to gluons an equal number
of gluinos $\la^a$. The gluinos affect the value of $b_0$ in Eq.
(\ref{cond}) but this seems to be the only change.
On the other hand, one might argue that $\langle (G_{\mu\nu}^a)^2\rangle$
now vanishes. Indeed the vacuum expectation 
value of the Lagrangian is zero \cite{voloshin}:
\be
\langle 0|-{1\over 4}(G_{\mu\nu}^a)^2+\bar{\la}^a
\slashchar{D}\la^a|0\rangle_{SUSY}~=~0\label{zero}
\ee
since it is an F-component of a superfield.
Since $\slashchar{D}\la^a=0$  
by virtue of equation of motion, one is inclined to think
that (\ref{zero}) implies the vanishing of the gluon condensate in
supersymmetric theories. Calculationally, it is not true, however,
and it is the vacuum expectation value
of the equation of motion, $\langle 0|\bar{\la}^a
\slashchar{D}\la^a|0\rangle_{SUSY}$, which is not vanishing 
in the renormalon approximation and cancels the gluon
condensate induced by renormalons. 
The reason is that in SUSY gluodynamics
the gluino wave function renormalization
is related to that of the gauge constant
while in ordinary QCD it is gauge dependent and in this sense
arbitrary. This might be an indication that
the dynamics of supersymmetric gauge theories is in fact very different
from QCD. 

In conclusion,
renormalons provide us with
with a systematic, although incomplete, way
to guess at non-perturbative physics in QCD. Theoretically,
there are important questions yet to be answered.

\section{PHENOMENOLOGY. GENERAL.}

\subsection{Could-be phenomenology.}

The main use of renormalons is in cases when there is no OPE.
However, one can try to gauge possible renormalon-based phenomenologies
to the case when an OPE is valid. 
As is mentioned above,
evaluating, for example, the T-product of two
electromagnetic currents at large Euclidean momenta $Q^2$
one finds that perturbation theory is unreliable in
the infrared as far as terms of order $Q^{-4}$ are concerned.
Then the polarization operator $\Pi (Q^2)$ could be represented
as
\bea
Q^2{d\Pi (Q^2)\over dQ^2}=~(parton~model)\cdot~~~~~~~~~~~~~~~\\ \nonumber
(1+a_1\aQ+a_2\aQ^2+...+
a_{ren}C{\La^4\over Q^4})\label{wb}
\eea
where $C$ is a constant related to the procedure of defining 
the uncertainty associated with an asymptotical expansion
and $a_{ren}$ varies with the choice of the external current, i.e. is
channel dependent. Fitting the data in various channel with a single unknown
constant $C$ one could try to develop a phenomenology
similar to QCD sum rules \cite{svz}.

This kind of phenomenology, however, would run into apparent difficulties.
Indeed, the $1/Q^4$ contribution in (\ref{wb}) is a tiny piece on
the background of the first terms in a 
$\aQ$ expansion. In particular any
redefinition of the coupling would reshuffle the whole series and
the $1/Q^4$ piece could well depend on such a redefinition.
Thus, it is inconsistent, generally speaking, to keep
the renormalon contributions without keeping many orders in
$\aQ$. The phenomenology is painstaking and its 
principal features are outlined in Ref.
\cite{mueller2,ms1}. For the implementation on the
lattice of this approach see \cite{ms}. 

On the other hand the success of the QCD sum rules is
based on a simplifying assumption that the non-perturbative
terms matching the renormalon 
 ambiguity are in fact large. 
It is natural to accept
this approach in other applications of renormalons
when we have no OPE, as well
\cite{az}.

\subsection{Renormalons and power corrections.}

Recent considerations of renormalons have
brought to light various  power corrections.
Common to all the examples which we list below is that 
they go beyond higher twist effects indicated
by the standard OPE.

({\it i}) In case of total cross section, a new type of correction
appears due to UV renormalons \cite{vz}. 
\be
{\sigma_{tot}(\gamma^{*}
\ra X)\over\sigma_{tot,parton}(\gamma^{*}\ra X)}
=1+a_1\as+...+C_{UV}{\La^2\over Q^2} 
.\ee
Combined with the idea of enhancement these terms could
solve certain problems with QCD 
sum rules and provide a link to NJL models \cite{vz2}.
We will concentrate
on IR renormalons and in this review
only note in passing that 
already the consideration of
the UV induced $\La^2/Q^2$ terms revealed the problem that overshadows
all the applications of renormalons. Namely, in the absence of OPE it
is much more difficult to relate different channels.
In particular, the $\La^2/Q^2$ corrections are welcome on
phenomenological grounds in the $\pi$-meson channel but not in
the $\rho$-meson channel. It is not known whether UV
renormalons produce such a pattern of $1/Q^2$ corrections. 

({\it ii}) Infrared renormalons induce
$1/Q$ corrections to many observables. The first
indications to these corrections 
were found in the cross section of the Drell-Yan process \cite{conto}
\be
h_1+h_2~\ra~(\mu^+\mu^-)+X\label{dy}
.\ee   
Shape variables, like the thrust $T$, also receive $1/Q$ corrections.
In the language of a finite gluon
 mass \cite{webber}:
\be
 1-T~\sim~\la/Q
.\ee
In all the cases these corrections are due to soft gluons
with 3-momenta of order $\la$. In the easiest way they can be visualized
on the example of a heavy quark mass \cite{bigi,bb}.
The infrared correction to a heavy mass $M_H$ due to the Coulomb-like field
is of order:
\be
{(\delta M_H)_{IR}\over M_H}~\sim~{1\over 8\pi M_H}\int_{\scriptstyle{IR}}
|{\bf E}^a|^2d^3{\bf r}~\sim~\as {\la \over M_H}\label{irm}\label{cc}
\ee
where ${\bf E}^a$ is the electric field and
by the infrared sensitive piece of the mass one can understand
the difference in mass renormalization in cases $\la=0$ and$\la\neq 0$.
This contribution is well defined then.

({\it iii}) Renormalons may bring new predictions also in cases
when the power corrections could be treated within the standard
OPE, like deep-inelastic scattering or inclusive decays
of heavy particles. The reason is that 
in terms of the standard procedures the renormalon calculus unifies 
evaluation of the coefficient functions and of the 
corresponding matrix elements.
As a result new relations may arise.
The simplest relation of this kind has been in fact already mentioned.
Namely, there is no dependence on the target.
We shall discuss further examples in the next section.

Thus, we conclude this section
with a remark that at least potentially renormalons 
may provide us with a new dimension in studies of power corrections.
These corrections, in turn, may be important, for
example, for the extraction of numerical values of $\as$.
from measurements of event shape variables. For an initial
attempt see \cite{hamacher}.

\section{ RENORMALON "ZEROS".}

\subsection{Heavy quark decays.}

Renormalon-based predictions naturally fall into two categories,
namely, when one gets either a vanishing or a nonvanishing
contribution. If we get a zero 
in a particular calculation, then it is natural to look
for a kind of more general explanation, like a symmetry.
This indeed turns out to be true, at least for the examples known
so far.

As a first example consider inclusive leptonic decays of heavy particles
\cite{bigi,bbz}. Confining ourselves to one-loop radiative corrections
and keeping $\la\neq 0$ we can, generally speaking, parametrize 
the infrared sensitivity in terms of the coefficients $a_i$
(see also Eq (\ref{par})):
\be
\Gamma_{tot}=\Gamma_{tot}^0(1+\al a_0ln\la^2+\al a_1\sqrt{\la^2}+...) 
\ee
where $\Gamma_{tot}^0$ is the partonic width, with inclusion of corrections
of order $\al$.
The results of a straightforward calculation are \cite{bbz}
\be
a_0~=~a_1~=~a_2~ =0.
\ee

Now, these zeros have different status in fact. The vanishing of
$a_0$ is the well-known Bloch-Nordsieck cancellation.
The vanishing of $a_1$ was claimed first \cite{bigi} 
on the basis of the OPE
for heavy quarks decays (for a review and further references see
\cite{vainshtein2}). This cancellation holds provided that the bare 
width is proportional to the fifth power of a short-distance mass
$M_{sh.d.}$ instead of the physical, or pole mass $M_{pole}$:
\be
M_{sh.d}~\approx~M_{pole}(1-{\as\over 2}\la)
\ee
where we keep only the infrared sensitive contribution.
The physical meaning of this procedure is simple. 
Indeed, the total decay width is sensitive to the
instanteneous energy release. 
The Coulomb field, on the other hand, is "shaken off" 
as a result of a fast decay
and the Coulomb correction to the mass (\ref{cc}) does not 
affect the total width. More elaborate calculations confirm this
intuition.

As for the vanishing of the coefficient $a_2$ there are no obvious
general reasons for it. Moreover, within the OPE one can show
\cite{vainshtein2} that the quadratic corrections are generally related
to matrix elements of operators $O_{1,2}$:
\be
O_1~=~{1\over M_H^2}\bar{Q}\sigma_{\mu\nu}G_{\mu\nu}Q, ~
O_2~=~{1\over M_H^2}\bar{Q}{\bf D}^2Q\label{oper}
\ee
where $Q$ is the (operator of) the field of the heavy particle,
$G_{\mu\nu}$ is the gluonic field strength tensor (with color indices
suppressed), and ${\bf D}$ is the covariant derivative. 
It is worth emphasizing that the use of OPE does not assume
that the matrix elements of the operators (\ref{oper})
over a free particle state are normalized to zero. Moreover,
the infrared-sensitive part of the matrix elements
are uniquely determined and are not subject to redefinitions.
It just happens that in the renormalon approximation 
the matrix elements of (\ref{oper}) vanish. This is an 
example of what we mean in point ({\it iii}) of the preceeding subsection
and we shall return to discuss it in more detail below.

\subsection{KLN-vacuum.}

In case of heavy quark decays reviewed above one 
expects the vanishing of the leading $1/Q$ corrections based on the OPE.
In case of the Drell-Yan process (\ref{dy}) there is no
OPE and one could expect appearance of $1/Q$ corrections.
However, a straightforward calculation demonstrated \cite{bb2}
that terms linear in $\la$ in fact cancel in one loop.
In more detail one evaluates moments $M_n$ from the cross section:
\be
\int d\tau\tau^{n-1}{d\sigma (Q^2,\tau)\over dQ^2}~=~M_n(1+\as a_1\sqrt{\la^2}+..)
\label{moments}\ee
where $Q$ is the invariant mass of the lepton
pair produced, $\tau=Q^2/s$ and $\sqrt{s}$ is 
the invariant mass of the $q\bar{q}$ from the initial 
hadrons $h_{1,2}$. The result \cite{bb2} is $a_1=0$
provided $n$ is not very large:
\be
n\cdot \Lambda_{QCD}/\sqrt{s}~\ll~1\label{ln}.
\ee

As argued in \cite{az2} the reason for this cancellation is
again general and it is a manifestation of the inclusive nature 
of the moments (\ref{moments}). If one considers, on the other
hand very large n (see (\ref{ln})) then the integral is practically
saturated by an exclusive channel. Moreover, the cancellation
of the linear terms in (at least $U(1)$) gauge theories
appears to be the same general phenomenon 
as the Bloch-Nordsieck cancellation.

One starts with the Kinoshita-Lee-Nauenberg 
theorem \cite{ln} as the most general statement on
infrared cancellations. Moreover, one can argue \cite{asz} that the
KLN summation over
initial and final states eliminates not
only the $ln\la$ terms as is emphasized in the original papers
but linear terms as well:
\be
\sum_{i,f}|S_{i\ra f}|^2~\sim~0\cdot ln\la^2+0\cdot\sqrt{\la^2}\label{canc}.
\ee
Here $S_{i\ra f}$ are elements of the $S$-matrix and relation
(\ref{canc}) holds in each order of the perturbative expansion.
The rationale behind (\ref{canc}) is simple:
the KLN summation cancels the singular, $1/\om$ terms on the
level of the amplitudes which implies elimination of both
$1/\om^2$ and $1/\om$ terms in $\sum |S_{i\ra f}|^2$.

Note that to visualize the cancellations due to the summation over
the degenerate initial states one may think in terms of a "KLN-vacuum"
which is populated by soft gluons.
To account for these particles in the initial state
the original KLN summation invokes both connected and disconnected graphs.
To prove Eqs. (\ref{canc}), (\ref{fold}) 
on the technical side it is crucial
that instead of summing over disconnected graphs one can systematically
add to ordinary Feynman graphs those with
propagators of soft particles changed into their complex
conjugates \cite{az2}:
\be
\left({-i\over k^2+i\epsilon}\right)~\ra~
\left({-i\over k^2+i\epsilon}\right)^{*}.\label{mod}
\ee
Adding graphs with the modified propagator (\ref{mod})
is equivalent to using the KLN vacuum and is simple technically.
It seems also plausible that the KLN vacuum
could be reduced to a finite-temperature vacuum but this
analogy has not been elaborated so far.

The next step is to relate the KLN sum, which extends over
initial and final states, into a summation over
the final states alone.
It is well known that as far as the most singular terms are concerned
it is indeed possible, and the KLN sum so to say folds into
twice the Bloch-Nordsieck sum over the final states:
\be
\sum_{\scriptstyle{i,f}}|S_{i\ra f}|^2~\ra_{\scriptstyle{soft}}2\cdot
\sum_{\scriptstyle{f}}|S_{i\ra f}|^2\label{fold}
\ee
where we have indicated that this is true for soft
but not collinear gluons. The new develpoment is to show
that Eq. (\ref{fold}) holds for linear terms as well.
The proof \cite{az2} utilizes the Low theorem and is made explicit for
the Drell-Yan process. However, the reasoning appears general enough
to apply to other processes as well . 

One may wonder also how far the use of the KLN vacuum 
(or, equivalently, of the propagator (\ref{mod})) extends infrared
cancellations.
The general answer \cite{asz} is that
the cancellations continue until one reaches the condensates terms.
In particular, in case of the gauge theories the
use of the propagator (\ref{mod})
doubles the effect of the perturbative gluon condensate (\ref{con}).
Very recently this function of the modified propagator (\ref{mod})
was emphasized in Ref. \cite{hoyer}.

To summarize: at the one-loop level, 
the linear terms cancel from inclusive 
cross sections the same way as logarithmic terms do.
The basic step in the proof is the use of the KLN vacuum
populated with soft particles or, equivalently, addition
of graphs with the modified propagator (\ref{mod}).
In case of $U(1)$ gauge theories the cancellation holds 
for higher loops as well.

\subsection{Vanishing martix elements.}

A specific feature of the renormalon calculus is that 
the power corrections get universally expressed 
in terms of $\La$ or $\la$ and are not dependent on the target.
On the other hand, if the same observable can be treated 
within OPE the power corrections are routinely related
to matrix elements of various operators. Thus, renormalons
fix the matrix elements. Whether this fixation provides
satisfactory results, is a different issue which has not
been addressed systematically, to our knowledge.
Thus, we confine ourselves to a few casual remarks.

As we have already mentioned, in case of heavy
quarks, renormalons imply the suppression 
\cite{bbz} of the matrix elements
of the operators (\ref{oper}):
\be
\langle free~particle|O_{1,2}|free~particle\rangle~=~O(\la^3)\label{lc}
\ee
while on dimensional grounds one would expect terms
of order $\la^2ln\la^2$.

Technically, the vanishing of the leading terms is due to
simple dynamical features of gauge interactions.
In particular one observes \cite{az3}
that the matrix element of the
operator of the kinetic 
energy, $\slashchar{{\bf D}}^2$ immediately reduces to a matrix element
of a local operator which is nothing else but the vacuum
expectation value
of the vector potential squared:
\be
\langle free~particle|\bar{Q}\slashchar{{\bf D}}^2
Q|free~particle\rangle
\sim C\cdot\langle{\bf A}^2\rangle\label{asq}
\ee
It is only natural then that the constant $C$ turns to be zero
because of gauge invariance. As for the matrix element
of the magnetic energy, $\bar{Q}\sigma_{\mu\nu}G_{\mu\nu}Q$,
its vanishing is due to the fact that transverse gluons do not
interact with a charged particle at rest.
While the the matrix elements in point, (\ref{lc}),
were calculated directly only at the one-loop level the reason for 
thier suppression remains true in higher orders as well \cite{az3}.

It is difficult to comment on the significance of (\ref{lc}).
On one hand, the theory of heavy quark decays (for a review see
\cite{vainshtein2}) assumes that the matrix elements in point
are determined by the atom-like structure of hadrons
and tacitly assumes that for free quarks they are zero.
The latter is not obvious (especially in case of confinement).
One may say then that this is supported by 
renormalons (see (\ref{lc})).
On the other hand the very idea that the matrix elements can be
target-independent looks very foreign to the whole OPE approach
to heavy hadrons decays. It appears more reasonable to apply
renormalons only to free particle decays.

In case of deep inelastic scattering one can evaluate power
corrections to moments of structure functions.
To be specific, consider \cite{mueller2}
the first moment of $F_3(x)$,
$\int dxF_3(x)$, relevant to the Llewellyn-Smith-Gross sum rule. 
Then the 
the leading twist contribution and the first power correction
are determined by the matrix elements of the 
following operator \cite{shuryak} :
\be
O_{\mu\nu}={2i \over q^2}\epsilon_{\mu\nu\alpha\beta}q_{\alpha}(
\bar{q}\gamma_{\beta}q+{4g\over9q^2}\bar{q}
\tilde{G}_{\beta\delta}\gamma_{\delta}\gamma_5q)\label{omn}
\ee
Applying the renormalon idea means that one evaluates
the power correction in terms of the matrix element of the
leading-twist operator. In terms of the IR parameters
entering the Feynman graphs this matrix element is a function
of the gluon mass $\la$, quark mass $m$ and of the quark virtuality
$p^2-m^2\equiv \epsilon^2$ \cite{az3}. In more detail:
\be
\langle  
{8g\over9}\bar{q}
\tilde{G}_{\alpha\beta}\gamma_{\beta}\gamma_5q\rangle=
f(p^2,m^2,\la^2){C_F \over 2\pi}{4\as\over 3}\langle 
\bar{q}\gamma_{\alpha}q\rangle
\ee
where $f(\la^2,m^2,\epsilon^2)$ is 
\bea
f(\la^2,m^2,\epsilon^2)=\int_{\scriptstyle{0}}^1dyX(y)lnX(y); \nonumber \\
X=\epsilon^2y(y-1)
+m^2y^2+\la^2(1-y).\label{ambig}
\eea
As one would expect the $\la^2ln\la^2$ term disappears
if $m^2\gg\la^2$ for the same reason as above (see Eq. (\ref{asq}))
and is taken over by the quark mass (for $\epsilon=0$) as an
infrared parameter. On the other hand, the $\la^2ln\la^2$
term does represent the power correction if other
infrared sensitive parameters are set to be zero.
In fact much more detailed calculations, representing 
the whole $x$-dependence of the quadratic power correction are
available in this case, or an equivalent thereof \cite{dmw,stein}.

At the next step one has to account for the anomalous dimension
of the operator governing the $Q^{-2}$ correction 
(see Eq. (\ref{omn})). 
In the Minkowskii-space approach the effect of the anomalous dimension
corresponds to emission
of soft gluons by energetic gluons. This has not been considered
so far and it is not clear that $\la\neq 0$ can be consistently
kept at this stage.
 
Summarizing this section, the
vanishing of certain power corrections revealed so far
through the use of renormalons 
can be understood each time within a broader
theoretical framework. The development of the corresponding
framework was sometimes initiated by renormalons and its completion
by including non-abelian theories still represents a challenge. 

\section{RENORMALONS AND EVENT SHAPES.}

Renormalon and renormalon-related techniques have turned out to be
instrumental in providing a theoretical basis for the
existence of $\La/Q$ corrections in shape variables
in  $e^+e^-$ annihilation. The phenomenology 
of these terms is of special interest
since they represent, on one hand, leading power
corrections and, on the other hand, there does not exist an
alternative more general framework to treat these
corrections. There are experimental fits to $1/Q$ corrections
\cite{barreiro,webber} and 
a careful experimental study of $1/Q$ has been made in \cite{hamacher}.

The very existence of the $1/Q$ corrections has been demonstrated
by various techniques. Let us mention finite gluon mass \cite{webber},
single renormalon chain \cite{az,ks}, dispersive approach to the 
running coupling \cite{dmw}.
It also can be seen from simple estimates. Consider, for example,
thrust T:
\be
T~=~max_{\scriptstyle{{\bf n}}}{\sum_{\scriptstyle{i}}|{\bf p_i
\cdot n}|\over \sum_{\scriptstyle{i}}|{\bf p}_i|}\label{thrust}
\ee
where ${\bf p}$ are the momenta of the particles produces 
while ${\bf n}$ is a unit vector. 
Perturbatively $T\neq 1$ arises because of the emission of
gluons from quarks. Consider then a contribution
to $T$ due to a soft gluon emission:
\be
\langle 1-T\rangle_{soft}\sim\int_0^{\La}
{\om\over Q}{d\om\over \om}\alpha_s(\La)\sim{\La\over Q}
\ee
where the first factor in the integrand comes from 
the definition of the thrust, $d\om/\om$
is the standard factor of emission of a soft gluon, and 
the running coupling $\alpha_s(\La)$ is of order unity.
Note that, unlike the inclusive Drell-Yan cross section,
evaluation of the thrust assumes that the momenta of final
particles are resolved on the infrared sensitive scale
and there is no reason, therefore, to expect cancellation of
these terms.

Once the existence of $1/Q$ is established, the effort
to create phenomenology shifts to deriving relations
among various observables and such relations were claimed
in all the approaches mentioned above.
In particular, in the one-renormalon approximation \cite{az}
one gets for the standard shape variables:
\bea
{1\over 2}\langle 1-T\rangle_{1/Q}={1\over 3\pi}\langle C\rangle_{1/Q}=
\\ \nonumber
={2\over \pi}\langle{\sigma_L\over \sigma_{T}}\rangle_{1/Q} =
{1\over \pi}\langle Esin^2\delta\rangle_{1/Q}=U\label{univ}
\eea
where $Q$ is now the total c.m. energy,
the subscript $1/Q$ means that only linear power corrections are kept
and $U$ is a universal factor:
\be
U~=~ {C_F\over \pi Q}\int_0^{\sim Q^2}{dk^2_{{\perp}}\over k_{{\perp}}}
\as (k^2_{\perp})~\sim~{\La\over Q}\label{u}
.\ee
Moreover, according to the rules of the renormalon calculus
only contribution of the Landau pole in $\as(k^2_{\perp})$, 
parametrized in a certain
way, is retained in (\ref{u}). As a result $U\sim\La/Q$ indeed.
Similar, although not identical, relations 
have been obtained within other approaches.
The earliest derivation \cite{webber} used 
the finite gluon mass technique. Comparisons with existing data,
in general,
look favourable \cite{webber,az,dmw}.

Having said this, we have to make 
numerous reservations as to the status
of relations of the type (\ref{univ}).
The point is that there are uncertainties in derivations
which can be removed only at a price of further assumptions.
In different aproaches these unertainties arise in different
ways but reflect the same difficulty: Namely, perturbative
calculations are reliable when the coupling is small.
Now we are trying to relate infrared contributions to various
observables. This is possible only if a certain
extrapolation procedure is accepted
and any procedure of this kind is speculative.

In the renormalon language, the problem is that
all orders of the perturbative expansion
which is an expansion in a small parameter in the UV region,
collapse to 
the same order of magnitude in the IR region.
Indeed, since
\be
\as^2(k^2_{\perp})~\sim~\La{d\as(k^2_{\perp})\over d\La}
\ee
we have
\be
\int_{\scriptstyle{IR}}
{dk^2_{\perp}\over k^2_{\perp}}\as^2(k^2_{\perp}){
k_{\perp}\over Q}~\sim~U~\sim~{\La\over Q}
.\ee
Thus, one is invited to address the problem in higher
orders as well.

There is a hope that the universality relations (\ref{u}) hold
in higher orders as well. Namely, it is known that all
the log terms which dominate in perturbative region are
universally related to the so called cusp anomalous dimension
$\gamma_{eik}$. If one retains only these terms in IR as well
then the universal factor $U$ in Eq (\ref{univ}) becomes
\cite{az,ks}:
\be
U~=~\int_0^{\sim Q^2}{dk^2_{{\perp}}\over Q\cdot k_{{\perp}}}
\gamma_{eik}(\as (k^2_{\perp})).
\ee
The reservation is that the terms which
dominate in UV region do not necessarily dominate upon
the continuation into the IR region.

An attractive possibility is to relate
the factor $U$ in Eq. (\ref{u})
to parameters of hadronization models \cite{az}. 
Indeed, the renormalon technique parametrizes contribution
of the region where the running coupling
$\as$ blows up. Since in the perturbative regime the coupling runs
with $k^2_{\perp}$ renormalons, at least intuitively, correspond
to introducing intrinsic transverse momentum for hadrons
in a quark jet. In the two-jet limit this relation can be 
made quantitative \cite{az}.
Namely, let $\tilde{\rho}(z,p_{\perp})$ 
denote the appropriately normalized distribution
of hadrons  
in a jet with longitudinal momentum fraction $z$ and  
perpendicular component $p_{\perp}$. Then
\be
U~\ra~\int d^2p_{\perp}\rho(p_{\perp}){p_{\perp}\over Q}\label{tube}
\ee
where $\rho(p_{\perp}\equiv\tilde{\rho}(0,p_{\perp}$.
Numerical value of (\ref{tube}) can be obtained from
fits to jet masses within the tube model (for a review see Ref.
\cite{webber3})
which identifies $\rho(p_{\perp})$ with the $p_{\perp}$ distribution
of hadrons in a rapidity-$p_{\perp}$ "tube".
Using the experimental data one then gets
\be
Q\cdot U~\approx~0.5 GeV
,\ee 
the value which also fits well the data on the 
$1/Q$ terms in shape variables.

Thus, Eq. (\ref{tube}) can be considered as an attempt to
formulate the enhancement hypothesis (see subsection 2.2) in pure
phenomenological terms. Theoretically it would be very attractive
to formulate this hypothesis in terms of matrix elements of some
operators. Note therefore the attempts to develop
a kind of OPE valid for jet physics \cite{ks}.

We have spelled out in some detail 
the difficulties of a phenomenology
based on renormalon chains. It is worthwhile to mention that
other approaches suffer uncertainties as well. For example,
the prediction for the thrust $T$ depends
on whether one keeps the gluon mass $\la\neq 0$ in the 
denominator of Eq. (\ref{thrust})
or not. The prediction closest to the renormalon chain arises
if this kinematical effect is neglected \cite{dmw}.

In view of the model dependence of the prediction for the
$1/Q$ corrections in shape variables, it would be important
to list predictions which could distinguish  between various
models. This has not been done however and we confine ourselves
only to a single remark of this kind \cite{az}.
Namely, the renormalon-chain predictions outlined above
allow easily for an enhancement hypothesis. That is, if
two-jet events are observed 
the $1/Q$ corrections to a heavy jet mass $M_h$
and to the light jet mass $M_l$ could be comparable. 
The only relation which is expected to hold is
\be
\langle 1-T\rangle~_{1/Q}=\langle{M_h^2\over Q^2}\rangle_{1/Q}+
\langle {M_l^2\over Q^2}\rangle_{1/Q}
.\ee
This relation is simply an expression of the fact
that the $1/Q$ corrections arise due to soft gluons.
On the other hand, the models with a finite gluon mass
or the frozen coupling
do not allow for such an enhancement.

Data at relatively low energies \cite{barreiro} do
indicate 
\be
\langle{M_h^2\over Q^2}\rangle_{1/Q}~\approx~
\langle {M_l^2\over Q^2}\rangle_{1/Q}
\ee
which can be considered as a support to the particular enhancement
mechanism described above.

To comprehend the significance of data at higher energies
more theoretical work is needed. The point is that
the $1/Q$ form of the leading power corrections has been
established in the two-jet limit.
At high energies, however, the two-jet events themselves are suppressed
by a Sudakov form-factor.
It is for this reason that the $1/Q$ corrections from the very
beginning \cite{conto} were claimed for resummed cross sections.
To ensure the two-jet dominance one could
introduce a corresponding weight factor. In case of the thrust,
for example one can consider \cite{ks} the following average
as far as the $1/Q$ terms are concerned:
\be
\langle 1-T\rangle_{1/Q}~\ra~\langle 
exp(-\nu (1-T))\rangle_{1/Q}
\ee
where $\nu$ is a new parameter which is to be large
enough
to ensure the dominance of the region $(1-T)\ll 1$.

To avoid 
a special weighting function one should have developed the theory
of $1/Q$ corrections for three-jet evens and so on. This has
not been done.  For a discussion of 
the effect of intrinsic $k_{\perp}$ near three-jet configurations
see Ref \cite{ellis}  

Summarizing this section,
relations among $1/Q$ terms in various observables are model
dependent. It looks plausible at this point 
that the renormalon-based
model will merge with the old-fashioned hadronization models.

\end{document}